\documentclass[aps,prd,onecolumn,amssymb,eqsecnum,nofootinbib]{revtex4}

\usepackage{epsf}  
\usepackage{graphicx}

\def\sun{\hbox{$\odot$}}

\begin{document}

% small logo
%\begin{figure*}
%\includegraphics[width = 0.10\textwidth]{caltech_logo}
%\end{figure*}

%%
%% NOTE: epsf package needed for this block
%%
%% CalTech Logo
%\hspace{-10mm}
%\leftline{\epsfbox{caltech_logo.eps}}
%\vspace{-10.0mm} % for revtex
%\thispagestyle{empty}
%{\baselineskip-4pt
%\font\fx=cmmib10 scaled\magstep2
%\leftline{\baselineskip20pt
%\hspace{25mm} % for revtex
%\vbox to0pt
%   { {\fx\hbox{California \hspace{1.5mm} Institute \hspace{1.5mm}
%of \hspace{1.5mm} Technology} }
%     {\large\sl\hbox{{TAPIR}} }\vss}}
%     
%\vspace{10mm}
% Logo end

%% UTB/CGWA Logo
%\hspace{-10mm}
%\leftline{
%{\epsfxsize=20mm \epsfysize=25mm 
%\epsfbox{UTB_logo.eps}}
%{\epsfxsize=30mm \epsfysize=20mm 
%\epsfbox{CGWA_logo.eps}}}
%\vspace{-10.0mm} % for revtex
%\thispagestyle{empty}
%{\baselineskip-4pt
%\font\fx=cmmib10 scaled\magstep2
%\leftline{\baselineskip20pt
%\hspace{50mm} % for revtex
%{\fx\hbox{University \hspace{1.5mm} of \hspace{1.5mm}
%Texas \hspace{1.5mm} at \hspace{1.5mm} Brownsville} }
%} \\ 
%\leftline{\baselineskip20pt
%\hspace{50mm} % for revtex
%{\fx\hbox{Center \hspace{1.5mm} for \hspace{1.5mm}
%Gravitational \hspace{1.5mm} Wave \hspace{1.5mm} Astronomy} }
%}}
%\vspace{5mm}
%% Logo end

%%preprint number
%\begin{flushright}
%
%\end{flushright}
%preprint number end

%\bigskip
%\bigskip
%\bigskip

%\begin{center}

\title{Modulation of the gravitational waveform 
by the effect of radiation reaction}

\author{Yasushi Mino}
\email{mino@tapir.caltech.edu}
\affiliation{mail code 130-33, 
California Institute of Technology, Pasadena, CA 91125}

\begin{abstract} 
When we calculate gravitational waveforms 
from extreme-mass-ratio inspirals (EMRIs) 
by metric perturbation, 
it is a common strategy to use the adiabatic approximation. 
Under that approximation, 
we first calculate the linear metric perturbation 
induced by geodesics orbiting  a black hole, 
then we calculate the adiabatic evolution 
of the parameters of geodesics due to the radiation reaction effect 
through the calculation of the self-force. 
This procedure is considered to be reasonable, 
however, there is no direct proof 
that it can actually produce the correct waveform 
we would observe. 
In this paper, we study the formal expression 
of the second order metric perturbation 
and show that it be expressed 
as the linear metric perturbation modulated 
by the adiabatic evolution of the geodesic. 
This evidence supports 
the assumption that the adiabatic approximation 
can produce the correct waveform, 
and that the adiabatic expansion we propose in Ref.\cite{adi} 
is an appropriate perturbation expansion 
for studying the radiation reaction effect on the gravitational waveform. 
\end{abstract}

\maketitle

%%%%%%%%%%%%%%%%%%%%%%%%%%%%%%%%%%%%%%%%%%%%%%%%%%%%%%%%%%%%
\section{Introduction}
%%%%%%%%%%%%%%%%%%%%%%%%%%%%%%%%%%%%%%%%%%%%%%%%%%%%%%%%%%%%

We study the gravitational waveforms 
from particles moving around Kerr black holes 
by using a metric perturbation method. 
There is an established method for calculating these waveforms 
from the linear metric perturbation of Kerr black holes. 
By the consistency of the Einstein equation, 
the source stress-energy tensor 
of the linear metric perturbation 
must satisfy the conservation law with respect to the background. 
As a result, the source of the linear metric perturbation 
moves along a geodesic of the background Kerr metric. 
Due to the integrability of the geodesic equation for the Kerr metric, 
its bound solutions are stable and have periodic features. 
To understand some feature of gravitational waves 
from such a stable system, 
we develop a technique of formal calculation\cite{rad}. 
The advantage of this technique is 
that, one can easily grasp some key features 
of the waveform without a complicated calculation, 
and it helps us to construct the strategy 
for an explicit calculation. 
Using this technique, we find 
that the waveform from the linear metric perturbation 
has a periodic feature\cite{rad} as we review in Sec.\ref{sec:lin}. 
Based on our understanding of this feature, 
a present numerical code is trying to identify 
which of the gravitational wave modes are 
strong enough to be observed 
by gravitational wave detectors\cite{drasco}. 

Because gravitational waves carry away 
energy and angular momentum, 
the system must have a dissipative evolution. 
This should change the periodic features 
of the gravitational waveform 
that we know from the linear metric perturbation. 
New features must be seen 
in the second order metric perturbation 
because the source term must include this dissipative effect. 
The goal of this paper is to find these new features. 
That is, the goal is to find 
how the wave amplitude and phase are modulated 
by gravitational radiation reaction. 
However, a serious calculation 
of the second order metric perturbation 
still has a lot of technical difficulties. 
We therefore extend the technique of formal calculation 
we developed to study the linear metric perturbation. 
We expect that the result of this formal calculation 
will be useful in future explicit calculations 
of the second order metric perturbation. 

This problem has attracted 
the attention of the gravity community\cite{capra} 
because its relevance 
to gravitational wave detectors, especially LISA. 
Among the primary targets for LISA are 
extreme-mass ratio inspirals (EMRIs), 
the inspiralling binary systems 
of supermassive black holes (with mass $\sim 10^5-10^9 M_{\sun}$) 
and stellar mass compact objects (with mass $\sim 1-100 M_{\sun}$). 
Because of the extreme mass ratios, 
the metric perturbation of the black hole 
is effective for studying the dynamics of the system; 
we use the Kerr geometry of the supermassive black hole as a background, 
and approximate the compact object as a point particle. 
The linear metric perturbation may predict the waveform at an instant. 
However, LISA will detect gravitational waves for several years. 
This time scale is comparable 
to the radiation reaction time scale of EMRIs 
and is essential to consider the dissipation effect 
of the gravitational radiation reaction 
because the effect will accumulate during the observation time. 

Most of the relevant investigations made so far 
discuss the calculation of the self-force \cite{sf} 
which needs regularization \cite{reg}. 
The underlying idea for such calculations is that: 
(1) At each instant, the orbit is approximated by a geodesic 
since, due to the extreme mass ratio, 
gravitational radiation reaction is a small effect. 
(2) As the effect of radiation reaction accumulates over time, 
the orbit changes from one geodesic to another. 
(3) These changes can be deduced from the self-force. 
This idea is usually referred as the `adiabatic' approximation. 
This calculation strategy has some theoretical problems. 
Since the self-force is gauge dependent, 
we may not be able to make a unique prediction 
about the orbital evolution. 
In Ref. \cite{adi}, we showed that, for times over which 
the standard metric perturbation expansion is valid, 
the self-force can be arbitrarily adjusted via gauge changes. 
As an extreme example, we showed that 
there is a gauge transformation 
which completely eliminates the self-force. 

Do we still need to calculate the self-force? 
In Ref. \cite{adi}, we argue that the answer may be yes. 
Under a certain gauge condition, 
the self-force may include the correct radiation reaction effect 
With this gauge condition, 
the calculation by the `adiabatic' approximation 
will give us the correct prediction of the gravitational waveform. 
The present paper discusses the self-force 
under that `physically reasonable' class of gauge conditions 
proposed in Ref. \cite{adi,rad} 
in the context of the second order metric perturbation. 
We will focus on gauge invariant quantities. 
The basic idea is that, 
because the gravitational waveform is observable, 
the features we can see from the waveform 
have invariant meanings. 
We will show that the self-force correctly describes 
the radiation reaction effect 
through the second order metric perturbation. 

The field equations 
for the linear and second order metric perturbations 
are derived as follows. 
We expand the metric in a small parameter $\epsilon$ 
\begin{eqnarray}
g_{\mu\nu} &=& g^{(0)}_{\mu\nu} +\epsilon g^{(1)}_{\mu\nu} 
+\epsilon^2 g^{(2)}_{\mu\nu} +O(\epsilon^3) \,, \label{eq:mpe} 
\end{eqnarray}
where $g^{(0)}_{\mu\nu}$ is the background metric. 
We insert the $\epsilon$-expansion of the metric 
into the Einstein tensor 
and expand it in powers of $\epsilon$. 
We formally obtain 
\begin{eqnarray}
G^{\mu\nu} &=& \epsilon G^{[1]\mu\nu}[{\bf g}^{(1)}] 
+\epsilon^2 \left\{G^{[1]\mu\nu}[{\bf g}^{(2)}] 
+G^{[2]\mu\nu}[{\bf g}^{(1)},{\bf g}^{(1)}]\right\} 
+O(\epsilon^3) \,, 
\end{eqnarray}
where $G^{[1]\mu\nu}[{\bf h}]$ is linear in $h_{\mu\nu}$ 
and $G^{[2]\mu\nu}[{\bf h}^{(1)},{\bf h}^{(2)}]$ is bi-linear 
in $h^{[1]\mu\nu}$ and $h^{(2)}_{\mu\nu}$. 
By similarity expanding the source term 
\begin{eqnarray}
T^{\mu\nu} &=& \epsilon T^{[1]\mu\nu}+\epsilon^2 T^{[2]\mu\nu}
+O(\epsilon^3) \,, 
\end{eqnarray}
we can formally write the perturbed Einstein equations 
to second order $\epsilon$ 
\begin{eqnarray}
G^{[1]\mu\nu}[{\bf g}^{(1)}] &=& 8\pi T^{[1]\mu\nu} 
\,, \label{eq:1ein} \\ 
G^{[1]\mu\nu}[{\bf g}^{(2)}] &=& 
-G^{[2]\mu\nu}[{\bf g}^{(1)},{\bf g}^{(1)}] 
+8\pi T^{[2]\mu\nu} \,. \label{eq:2ein} 
\end{eqnarray}

In Sec.\ref{sec:lin}, we review the derivation of (\ref{eq:1ein}) \cite{rad}. 
Due to the consistency of the Einstein equation, 
the source of the linear metric perturbation, $T^{[1]\mu\nu}$, 
must be a geodesic. 
Because we are interested in a bound geodesic, 
bound geodesics are triperiodic, 
and we will see that the linear metric perturbation 
induced by it is as well. 
Before discussing the second order equation (\ref{eq:2ein}), 
we make a formal argument about the self-force 
and the orbital evolution due to this effect in Sec. \ref{sec:lsf} 
because those effects should be included 
in the source of the second order equation, $T^{[2]\mu\nu}$. 
In Sec. \ref{sec:2nd}, we discuss the formal analysis 
of the second order Einstein equation (\ref{eq:2ein}). 
The second order metric perturbation is induced 
by the quadratic term of the linear metric perturbation, 
and by the second order source term. 
We emphasize these two effects leads 
to substantially different features 
of the second order metric perturbation. 
Sec. \ref{sec:con} concludes our result. 

Our background geometry will be a Kerr black hole 
with mass $M$ and spin parameter $a$. 
We will a point particle with mass $\mu$ 
as the source of the metric perturbation. 
Throughout this paper, we use Boyer-Lindquist coordinates, 
$\{t,r,\theta,\phi\}=\{x^\alpha\}$. 
We adopt the geometrized units which are defined such that $G=c=1$.

%%%%%%%%%%%%%%%%%%%%%%%%%%%%%%%%%%%%%%%%%%%%%%%%%%%%%%%%%%%%
\section{Linear Metric Perturbation}
\label{sec:lin}
%%%%%%%%%%%%%%%%%%%%%%%%%%%%%%%%%%%%%%%%%%%%%%%%%%%%%%%%%%%%

In this section, 
we review the periodic feature of the linear metric perturbation. 
It is well known that, when the background is a vacuum solution, 
the source term of the linear metric perturbation 
must be conserved with respect to the background. 
Since we use a point mass source, 
this means that its world line is a geodesic of the background geometry. 
Because we are interested in gravitational waves 
from binaries in the inspiralling phase, 
we only consider bound geodesics. 

We denote the orbital coordinates of the geodesic 
by $\{\bar x^\alpha\}$ 
and the $4$-velocity by $\bar v^\alpha :=d\bar x^\alpha/d\tau$, 
where $\tau$ is the proper time. 
The geodesic equation for a Kerr black hole 
has three nontrivial constants of motion: 
the energy $E=\mu \eta^E_\alpha \bar v^\alpha$, 
the $z$-component of angular momentum 
$L=\mu \eta^L_\alpha \bar v^\alpha$ 
and the Carter constant 
$C=(\mu/2)\eta_{\alpha\beta}\bar  v^\alpha \bar v^\beta$. 
Here $\eta^E_\alpha$ and $\eta^L_\alpha$ are 
temporal and rotational Killing vectors, respectively, 
and $\eta_{\alpha\beta}$ is the Killing tensor. 
Hereafter we collectively denote these constants by ${\cal E}^a$. 
The geodesic equations can then be formally written as 
\begin{eqnarray}
\left({d\bar r \over d\lambda}\right)^2 &=& R(\bar r;{\cal E}^a) 
\,,\quad 
\left({d\bar\theta \over d\lambda}\right)^2 
= \Theta(\bar\theta;{\cal E}^a) 
\,, \label{eq:geo1} \\
{d\bar t \over d\lambda} &=& 
T_r(\bar r;{\cal E}^a)+T_\theta(\bar\theta;{\cal E}^a) 
\,,\quad 
{d\bar\phi \over d\lambda} 
= \Phi_r(\bar r;{\cal E}^a)+\Phi_\theta(\bar\theta;{\cal E}^a) 
\,, \label{eq:geo2} 
\end{eqnarray}
where $\lambda$ is the affine parameter of the orbit 
related with the proper time 
by $\lambda = \int d\tau/(\bar r^2+a^2 \cos^2\bar \theta)$. 
For bound geodesics, 
the $r$-motion and the $\theta$-motion are periodic 
and one can write them as a discrete Fourier series. 
We formally write 
\begin{eqnarray}
\bar r &=& \sum_k r_k e^{ik\chi_r} \,, \quad 
\chi_r = \Upsilon_r(\lambda -\lambda^r) 
\,, \label{eq:geo3} \\ 
\bar \theta &=& \sum_l \theta_l e^{il\chi_\theta} 
\,, \quad 
\chi_\theta = \Upsilon_\theta(\lambda -\lambda^\theta) 
\,, \label{eq:geo4} 
\end{eqnarray}
where $\lambda^r$ and $\lambda^\theta$ 
are constants for integration. 
The expansion coefficients, $r_k$ and $\theta_l$, 
and the effective frequencies, 
$\Upsilon_r$ and $\Upsilon_\theta$, 
are functions of the constants of motion ${\cal E}^a$. 
Using (\ref{eq:geo3}) and (\ref{eq:geo4}), 
we can formally integrate (\ref{eq:geo2}) as 
\begin{eqnarray}
\bar t &=& \chi_t
+\sum_k t^r_k e^{ik\chi_r}+\sum_l t^\theta_l e^{il\chi_\theta} 
\,, \quad 
\chi_t = \Upsilon_t(\lambda-\lambda^t)
\,, \label{eq:geo5} \\ 
\bar \phi &=& \chi_\phi 
+\sum_k \phi^r_k e^{ik\chi_r}+\sum_l \phi^\theta_l e^{il\chi_\theta} 
\,, \quad 
\chi_\phi = \Upsilon_\phi(\lambda-\lambda^\phi)
\,, \label{eq:geo6} 
\end{eqnarray}
where $\lambda^t$ and $\lambda^\phi$ 
are constants of integration. 
The expansion coefficients, 
$t^r_k$, $t^\theta_k$, $\phi^r_k$ and $\phi^\theta_l$, 
and the effective frequencies, 
$\Upsilon_t$ and $\Upsilon_\phi$, 
are functions of the constants of motion ${\cal E}^a$. 
In summary, one can specify a geodesic 
by three constants of motion ${\cal E}^a$ 
and four constants of integration $\lambda^\alpha$. 
Because one is free to choose 
the zero of the affine parameter $\lambda$, 
one of four integral constants $\lambda^\alpha$ 
is not physically significant, 
and the gravitational waveform induced by the geodesic 
depends only on three differences of the four integral constants 
[See (\ref{eq:gwf}).]. 

Under a certain gauge condition, one can define 
a tensor Green's function 
for the linear Einstein equation (\ref{eq:1ein}). 
Because the background is stationary and axisymmetric, 
this tensor Green's function has the form 
\begin{eqnarray}
G_{\alpha\beta\mu\nu}(x,x') 
= G_{\alpha\beta\mu\nu}(t-t',\phi-\phi';r,r',\theta,\theta') 
\,. \label{eq:tgr} 
\end{eqnarray}
For the linear metric perturbation, 
the source term is described 
by the stress-energy tensor of a point particle \cite{sf} 
written as 
\begin{eqnarray}
T^{[1]\mu\nu} &=& \mu \int d\tau \, 
\bar v^\mu \bar v^\nu {\delta(x-\bar x(\tau)) \over \sqrt{-|g^{[0]}|}} 
\,, \label{eq:st1} 
\end{eqnarray}
where $|g^{[0]}|$ is 
the determinant of the background metric. 
The corresponding linear metric perturbation is given by 
\begin{eqnarray}
g^{[1]}_{\alpha\beta}(x) 
= 8\pi\mu\int d\lambda \left({d\tau \over d\lambda}\right) 
G_{\alpha\beta\mu\nu}\left(x,\bar x(\lambda)\right)
\bar v^\mu \bar v^\nu
\,. \label{eq:met} 
\end{eqnarray}
Using the formal expression for the geodesic, 
(\ref{eq:geo3})-(\ref{eq:geo6}), 
we obtain the following formal expression 
for this linear metric perturbation 
\begin{eqnarray}
g^{[1]}_{\alpha\beta}(x) &=& \sum_{k,l,m}
e^{-i \omega_{(k,l,m)} t+i m \phi}
g^{[1]}_{\alpha\beta(k,l,m)}(r,\theta)
e^{i (k \omega_r t^r +l \omega_\theta t^\theta 
+m \omega_\phi t^\phi)} 
\,, \label{eq:lmp} \\ 
\omega_{(k,l,m)} &=& k \omega_r +l \omega_\theta +m \omega_\phi \,, \quad 
\omega_r = {\Upsilon_r \over \Upsilon_t} \,, \quad 
\omega_\theta = {\Upsilon_\theta \over \Upsilon_t} \,, \quad 
\omega_\phi = {\Upsilon_\phi \over \Upsilon_t} \,, \\ 
t^r &=& \Upsilon_t(\lambda^r-\lambda^t) \,, \quad 
t^\theta = \Upsilon_t(\lambda^\theta-\lambda^t) \,, \quad 
t^\phi = \Upsilon_t (\lambda^\phi-\lambda^t) 
\,, 
\end{eqnarray}
where the expansion coefficients, 
$g^{[1]}_{\alpha\beta(k,l,m)}(r,\theta)$, 
depend only on the constants of motion, ${\cal E}^a$, 
and are independent of $\lambda^r$, $\lambda^\theta$, 
$\lambda^t$ and $\lambda^\phi$. 

The Green's function (\ref{eq:tgr}) 
can be obtained under the Lorenz gauge or the Harmonic gauge. 
However, it can be defined 
under a larger class of gauge conditions 
which we call by the `physically reasonable class'. 
Its definition can be understood 
by considering the residual gauge transformation. 
By the gauge transformation $x^\alpha \to x^\alpha + \xi^\alpha$ 
with the gauge field of the form 
\begin{eqnarray}
\xi^\alpha(x) &=& \sum_{k,l,m}
e^{-i \omega_{(k,l,m)} t+i m \phi}
\xi^\alpha_{(k,l,m)}(r,\theta)
e^{i (k \omega_r t^r +l \omega_\theta t^\theta 
+m \omega_\phi t^\phi)} 
\,, 
\end{eqnarray}
the expansion coefficients $h_{\alpha\beta(k,l,m)}(r,\theta)$ 
are transformed, 
but, the formal expression 
of the linear metric perturbation (\ref{eq:lmp}) 
is invariant. 

The main features of the gravitational waveforms 
can be read off from the formal expression (\ref{eq:lmp}). 
For any distant observer at $r\to\infty$ 
and a specific angular position, 
the waveform can be written as 
\begin{eqnarray}
h(t) &=& \sum_{k,l,m}h_{(k,l,m)} 
e^{-ik\omega_r(t-t^r)-il\omega_\theta(t-t^\theta) 
-im\omega_\phi(t-t^\phi)} 
\,. \label{eq:gwf} 
\end{eqnarray}
Because the characteristic frequencies of waves 
$\omega_r$, $\omega_\theta$, and $\omega_\phi$ 
are all associated with geodesics, 
we may conclude 
that the waveform (\ref{eq:gwf}) does not reflect 
dissipative effects occurring on the radiation reaction time scale. 
This is to be expected 
since the source of the linear metric perturbation 
moves on a stable geodesic.

%%%%%%%%%%%%%%%%%%%%%%%%%%%%%%%%%%%%%%%%%%%%%%%%%%%%%%%%%%%%
\section{Self-Force and the Orbital evolution}
\label{sec:lsf}
%%%%%%%%%%%%%%%%%%%%%%%%%%%%%%%%%%%%%%%%%%%%%%%%%%%%%%%%%%%%

Because the gravitational wave part of the linear metric perturbation 
carries away energy, 
the orbit of the particle deviates away from a geodesic. 
The effect of the orbital deviation is described 
by the so-called self-force, 
that is both induced by and acting on the particle itself. 
Because the metric perturbation 
induced by the particle diverges along the orbit, 
a regularization prescription is necessary 
to obtain the finite self-force \cite{sf}. 
Up to leading order in the particle's mass, 
we developed the regularization prescription 
by the technique of the matched asymptotic expansion. 
The final result is now called the MiSaTaQuWa self-force \cite{sf}, 
and is formally written as 
\begin{eqnarray}
{D \over d\tau}v^\alpha &:=& 
\mu f^\alpha(\tau) = \lim_{x \to \bar x(\tau)} 
\mu f^\alpha[{\bf g}^{[1]}-{\bf g}^{[1]sing.}](x) 
\,, \label{eq:reg1} 
\end{eqnarray}
where the bare term ${\bf g}^{[1]}$ represents 
the full linear metric perturbation induced by the point particle, 
and the counter term ${\bf g}^{[1]sing.}$ is 
the singular part of the linear metric perturbation 
to be subtracted for regularization. 
$f^\alpha[]$ is a derivative operator for the self-force. 
Because both terms ${\bf g}^{[1]}$ and ${\bf g}^{[1]sing.}$ 
are divergent along the geodesic, 
it is necessary to evaluate them 
at a field point $x^\alpha$, not at the geodesic. 
After the subtraction, 
${\bf g}^{[1]}-{\bf g}^{[1]sing.}$ becomes regular along the geodesic, 
and we may take the limit $x^\alpha \to \bar x^\alpha$ 
to obtain the finite self-force. 

One can use the formal expression (\ref{eq:lmp}) 
to evaluate the bare term ${\bf g}^{[1]}$. 
The formal expression (\ref{eq:lmp}) is also applicable 
for the counter term ${\bf g}^{[1]sing.}$ 
because one can derive 
the Green's function for the counter term 
in the same form (\ref{eq:tgr}) 
as we discussed in Ref. \cite{rad}. 
After some formal calculations, 
the self-force can be written as 
\begin{eqnarray}
\mu f^\alpha &=& 
\sum_{k,l} \mu f^\alpha_{k,l} e^{ik\chi_r +il\chi_\theta}
\,. \label{eq:sf0} 
\end{eqnarray}

It is crucial to observe that the orbital equations 
(\ref{eq:geo1})and (\ref{eq:geo2}) 
are still true for time dependent ${\cal E}^a$, 
and we will use these equations 
to calculate the orbital evolution due to the self-force. 
Quantities that evolve due to the self-force 
are denoted with tilde. 
The evolution equations for $\tilde{\cal E}^a$ are then 
\begin{eqnarray}
{d \over d\lambda}\tilde E 
&=& \left({d\tau \over d\lambda} \right) 
\mu^2\eta^E_\alpha f^\alpha 
\,, \quad 
{d \over d\lambda}\tilde L 
= \left({d\tau \over d\lambda} \right) 
\mu^2\eta^L_\alpha f^\alpha 
\,, \quad 
{d \over d\lambda}\tilde C 
= \left({d\tau \over d\lambda} \right) 
\mu^2\eta_{\alpha\beta} \bar v^\alpha f^\beta 
\,. \label{eq:sf1} 
\end{eqnarray}
If each of the $\tilde {\cal E}^a$ is expanded 
in a manner similar to (\ref{eq:sf0}). 
these equations can be formally integrated to give 
\begin{eqnarray}
\tilde E &=& E_0 +\left<\dot E\right>\lambda 
+\sum_{k,l} E_{k,l} e^{ik\chi_r +il\chi_\theta} 
\,, \quad 
\tilde L = L_0 +\left<\dot L\right>\lambda 
+\sum_{k,l} L_{k,l} e^{ik\chi_r +il\chi_\theta} 
\,, \label{eq:sf2} \\
\tilde C &=& C_0 +\left<\dot C\right>\lambda 
+\sum_{k,l} C_{k,l} e^{ik\chi_r +il\chi_\theta} 
\,, \label{eq:sf3} 
\end{eqnarray}
where ${\cal E}^a_0$ denote the initial values at $\lambda=0$
\footnote{We choose ${\cal E}^a_{0,0}$, such that 
$\sum_{k,l} {\cal E}^a_{k,l} e^{ik\chi_r +il\chi_\theta}=0$ 
when $\lambda=0$.}. 
$\left<\dot {\cal E}^a\right>$ and ${\cal E}^a_{k,l}$ 
are of order $(\mu/M)^2$ and due to the self-force. 

The orbital evolution can be derived. 
from (\ref{eq:geo1}), (\ref{eq:geo2}), (\ref{eq:sf2}) and (\ref{eq:sf3}). 
Following our notation convention introduced above, 
the inspiralling world line is denoted by $\tilde x^\alpha$. 
We define the $r$- and $\theta$-motions by 
\begin{eqnarray}
\tilde r &=& \sum_k \tilde r_k e^{ik\tilde\chi_r} 
\,, \quad 
\tilde \theta = \sum_l \tilde\theta_l e^{il\tilde\chi_\theta} 
\,, \label{eq:sf4} 
\end{eqnarray}
where the expansion coefficients 
$\tilde r_k$ and $\tilde\theta_l$ 
are the same as those in (\ref{eq:geo3}) and (\ref{eq:geo4}), 
but they are functions of $\tilde{\cal E}^a$ 
instead of ${\cal E}^a$. 
The evolution equations 
for $\tilde\chi_r$ and $\tilde\chi_\theta$ are given by 
\begin{eqnarray}
{d \over d\lambda}\tilde\chi_r &=& \tilde\Upsilon_r
-{\sum_k (d_\lambda\tilde r_k) e^{ik\tilde\chi_r} 
\over \sum_{k'}ik' \tilde r_{k'} e^{ik'\tilde\chi_r}}
\,, \quad 
{d \over d\lambda}\tilde\chi_\theta = \tilde\Upsilon_\theta
-{\sum_l (d_\lambda\tilde\theta_l) e^{il\tilde\chi_\theta} 
\over \sum_{l'}il' \tilde\theta_{l'} e^{il'\tilde\chi_\theta}}
\,, \label{eq:sf5} 
\end{eqnarray}
where $d_\lambda=(d/d\lambda)$ acts 
on $\tilde{\cal E}^a$ of $\tilde r_k$ and $\tilde \theta_l$. 
The effective frequencies 
$\tilde\Upsilon_r$ and $\tilde\Upsilon_\theta$ 
are the same as $\Upsilon_r$ and $\Upsilon_\theta$ 
but are functions of $\tilde{\cal E}^a$ 
instead of ${\cal E}^a$. 
Similarly, we define the $t$- and $\phi$-motions by 
\begin{eqnarray}
\tilde t &=& \tilde\chi_t
+\sum_k \tilde t^r_k e^{ik\tilde\chi_r}
+\sum_l \tilde t^\theta_l e^{il\chi_\theta} 
\,, \quad 
\tilde \phi = \tilde\chi_\phi 
+\sum_k \tilde\phi^r_k e^{ik\chi_r}
+\sum_l \tilde\phi^\theta_l e^{il\chi_\theta} 
\,. \label{eq:sf6} 
\end{eqnarray}
Then, the evolution equations 
for $\tilde\chi_t$ and $\tilde\chi_\phi$ are 
\begin{eqnarray}
{d \over d\lambda}\tilde\chi_t &=& \tilde\Upsilon_t 
+\sum_k \left\{
ik\left(d_\lambda\tilde\chi_r -\tilde\Upsilon_r\right)
\tilde t^r_k-d_\lambda\tilde t^r_k \right\} 
e^{ik\tilde\chi_r}
\nonumber \\ && \qquad 
+\sum_l \left\{
il\left(d_\lambda\tilde\chi_\theta-\tilde\Upsilon_\theta\right)
\tilde t^\theta_l-d_\lambda \tilde t^\theta_l \right\} 
e^{il\tilde\chi_\theta}
\,, \label{eq:sf7} \\ 
{d \over d\lambda}\tilde\chi_\phi &=& \tilde\Upsilon_\phi 
+\sum_k \left\{
ik\left(d_\lambda\tilde\chi_r -\tilde\Upsilon_r\right)
\tilde\phi^r_k -d_\lambda \tilde\phi^r_k\right\} 
e^{ik\tilde\chi_r}
\nonumber \\ && \qquad 
+\sum_l \left\{
il\left(d_\lambda\tilde\chi_\theta -\tilde\Upsilon_\theta\right)
\tilde\phi^\theta_l-d_\lambda\tilde\phi^\theta_l \right\} 
e^{il\tilde\chi_\theta}
\,, \label{eq:sf8} 
\end{eqnarray}
where the effective frequencies 
$\tilde\Upsilon_t$ and $\tilde\Upsilon_\phi$ 
are the same as $\Upsilon_t$ and $\Upsilon_\phi$ 
but are functions of $\tilde{\cal E}^a$ instead of ${\cal E}^a$. 

We now derive the formal expression of $\tilde\chi_\alpha$ 
by perturbation. 
Recall that $\tilde\chi_\alpha-\chi_\alpha$ is 
of order $\mu/M$ by definition. 
The formal expressions (\ref{eq:sf5}), (\ref{eq:sf7}) and (\ref{eq:sf8}) 
can therefore be written in the form 
\begin{eqnarray}
{d \over d\lambda}\tilde\chi_\alpha &=& 
\Upsilon_{\alpha(0)}+\left<\dot \Upsilon_\alpha\right>\lambda 
+\sum_{k,l} \Upsilon_{\alpha(k,l)}e^{ik\chi_r+il\chi_\theta} 
\,, 
\end{eqnarray}
where $\Upsilon_{\alpha(0)}=\Upsilon_\alpha|_{{\cal E}={\cal E}_0}$ 
is for the background orbital evolution. 
The formal expression of $\tilde\chi_\alpha$ becomes 
\begin{eqnarray}
\tilde\chi_\alpha &=& 
\Upsilon_{\alpha(0)} (\lambda-\lambda^\alpha)
+{1 \over 2}\left<\dot \Upsilon_\alpha\right>\lambda^2 
+\Upsilon_{\alpha(0,0)}\lambda 
+\sum_{k,l} \tilde\chi_{\alpha(k,l)}e^{ik\chi_r+il\chi_\theta} 
\,, \label{eq:sf9} 
\end{eqnarray}
where we set 
such that $\tilde\chi_\alpha=\chi_\alpha$ when $\lambda=0$
\footnote{This is achieved by choosing $\tilde\chi_{\alpha(0,0)}$
such that $\sum_{k,l} \tilde\chi_{\alpha(k,l)}
e^{ik\chi_r+il\chi_\theta}=0$ at $\lambda=0$.}. 
This expression has two key features. 
One is the linear growth from 
${1 \over 2}\left<\dot \Upsilon_\alpha\right>\lambda^2$. 
This effect comes from the linear growth of ${\cal E}^a$ 
since $<\dot \Upsilon_\alpha>=
(\partial \Upsilon_\alpha/\partial {\cal E}^a)<\dot {\cal E}^a>$. 
This is expected to give the dominant phase evolution 
of the gravitational waveform 
because of the quadratic growth of the phase. 
It is the reason that the orbit's phase 
deviates from that of a geodesic 
on the dephasing time scale $(\propto 1/\sqrt{\mu})$, 
while its frequencies deviate 
on the radiation reaction time scale $(\propto 1/\mu$). 
The second feature is a small shift 
of the time averaged frequencies by $\Upsilon_{\alpha(0,0)}$. 
The phase evolution due to this effect 
will accumulate in time \cite{pp}, 
however it remains small over the radiation reaction time 
and is therefore not likely to be observable \cite{adi}.

%%%%%%%%%%%%%%%%%%%%%%%%%%%%%%%%%%%%%%%%%%%%%%%%%%%%%%%%%%%%
\section{The Second Order Metric Perturbation}
\label{sec:2nd} 
%%%%%%%%%%%%%%%%%%%%%%%%%%%%%%%%%%%%%%%%%%%%%%%%%%%%%%%%%%%%

In this section we derive formal solutions 
of the second order Einstein equation (\ref{eq:2ein}). 
By using the same gauge condition as we solve (\ref{eq:1ein}), 
we can formally integrate (\ref{eq:2ein}) 
with the tensor Green's function (\ref{eq:tgr}). 
We separate the second order metric perturbation 
into pieces 
\begin{eqnarray}
g^{[2]}_{\alpha\beta} &=& 
g^{[2]NL}_{\alpha\beta}+g^{[2]SF}_{\alpha\beta} 
\,, 
\end{eqnarray} 
where $g^{[2]NL}_{\alpha\beta}$ is 
due to the nonlinearity of the Einstein equation 
and $g^{[2]SF}_{\alpha\beta}$ is 
from the perturbation of the source. 
(\ref{eq:2ein}) then becomes 
\begin{eqnarray} 
G^{[1]\mu\nu}[{\bf g}^{(2)NL}] &=& 
-G^{[2]\mu\nu}[{\bf g}^{(1)},{\bf g}^{(1)}] 
\,, \quad 
G^{[1]\mu\nu}[{\bf g}^{(2)SF}] = 
8\pi T^{[2]\mu\nu} 
\,. 
\end{eqnarray}

We first discuss the formal calculation of ${\bf g}^{[2]NL}$. 
By the Green's method, we have 
\begin{eqnarray} 
g^{[2]NL}{}_{\alpha\beta}(x) &=& -\int_{-\infty<r^*<\infty} 
d x'^4\, G_{\alpha\beta\mu\nu}
G^{[2]\mu\nu}[{\bf g}^{(1)},{\bf g}^{(1)}] 
\,, \label{eq:2mpNL} 
\end{eqnarray}
where the domain of radial integration is 
outside the outer horizon and inside the future infinity; 
with $r^*=\int dr(r^2+a^2)/(r^2-2Mr+a^2)$, 
this is defined as $-\infty<r^*<\infty$.
Because the Kerr geometry is stationary an axisymmetric, 
the differential operator 
$G^{[2]\mu\nu}[{\bf h},{\bf k}]$ 
does not include terms that explicitly depend on $t$ and $\phi$. 
Therefore, the formal expression of the source term is 
\begin{eqnarray}
G^{[2]\mu\nu}[{\bf g}^{(1)},{\bf g}^{(1)}] &=& \sum_{k,l,m}
e^{-i \omega_{(k,l,m)} t+i m \phi}
G^{[2]\mu\nu}_{(k,l,m)}(r,\theta) 
e^{i (k \omega_r t^r +l \omega_\theta t^\theta 
+m \omega_\phi t^\phi)} 
\,, \label{eq:2ein0} 
\end{eqnarray}
where we have used (\ref{eq:lmp}). 
Using the Green's function of the form (\ref{eq:tgr}), 
one may obtain the formal expression 
\begin{eqnarray}
g^{[2]NL}{}_{\alpha\beta}(x) &=& \sum_{k,l,m}
e^{-i \omega_{(k,l,m)} t+i m \phi}
g^{[2]\alpha\beta}_{(k,l,m)}(r,\theta)
e^{i (k \omega_r t^r +l \omega_\theta t^\theta 
+m \omega_\phi t^\phi)} 
\,. \label{eq:2mpNL0} 
\end{eqnarray}
We see that ${\bf g}^{[2]NL}$ has the same formal expression 
as the first order term (\ref{eq:lmp}). 
This part of the second order metric perturbation 
changes on the dynamical time scale of the orbit 
which is much shorter than the radiation reaction time scale. 
This means that ${\bf g}^{[2]NL}$ does not include 
radiation reaction effects, 
and that ${\bf g}^{[2]NL}$ describes 
the non-linear effect of the Einstein equation. 
This effect makes a small correction to the wave amplitude 
of the linear metric perturbation, (\ref{eq:lmp}), 
and is not likely to be observable. 

Although this conclusion may look reasonable, 
this argument is not yet well-supported mathematically 
because (\ref{eq:2mpNL}) is actually divergent. 
In order to evaluate the formal expression of ${\bf g}^{[2]NL}$, 
it is necessary to see the possible change 
due to the regularization calculation. 
The integration of (\ref{eq:2mpNL}) has two kinds of divergences. 
One was pointed out in Ref. \cite{2nd}, 
where it was referred to as an ultraviolet divergence. 
Near the particle, 
the singular behavior of the linear metric perturbation 
under the harmonic gauge condition is 
${\bf g}^{[1]} \propto \mu/R$, 
where $R$ is the spatial distance from the particle 
in the local inertia frame. 
Because the Einstein tensor is a second order differential operator, 
we have 
$G^{[2]\mu\nu}[{\bf g}^{(1)},{\bf g}^{(1)}] \propto \mu^2/R^4$. 
The Green's function behaves as $\propto 1/R$. 
Therefore the spatial integral in (\ref{eq:2mpNL}) 
diverges as $1/R^2$ near the orbit. 
The other kind of divergence, 
an infrared divergence in the field theory terminology, 
is due to the fact that the graviton is a massless particle. 
Because the geodesic source of the linear metric perturbation 
is a stable orbit, 
it can radiate an infinite amount of energy 
from the infinite past, 
which is stored in a Cauchy surface. 
This becomes a source for the second order perturbation 
given by (\ref{eq:2mpNL}). 
At spatial infinity, the linear metric perturbation 
of a specific frequency under the harmonic gauge condition 
behaves asymptotically as 
${\bf g}^{[1]} \sim \mu e^{-i \omega(t-r)}/r$ 
where $r$ is the radius in asymptotically flat coordinates. 
This means that 
$G^{[2]\mu\nu}[{\bf g}^{(1)},{\bf g}^{(1)}] 
\sim \mu^2 \omega\omega' e^{-i (\omega+\omega')(t-r)}/r^2$, 
and the spatial integral of (\ref{eq:2mpNL}) diverges 
as $\ln(r)$ at infinity. 
In App. \ref{app:reg1} and App. \ref{app:reg2}, we introduce 
the regularization prescription for (\ref{eq:2mpNL}) 
and show that it does not alter the result. 
By applying a gauge transformation, 
the result should apply for metric perturbations 
under broader gauge conditions. 

We next discuss the formal calculation of ${\bf g}^{[2]SF}$. 
In order to identify the stress-energy tensor 
for the second order metric perturbation, 
we recall the derivation of the self-force 
by the linear metric perturbation. 
In Ref. \cite{sf}, the self-force was derived 
using a matched asymptotic expansion 
of the black hole metric for the particle 
and the linear metric perturbation away from the orbit. 
The singular part of the linear metric perturbation 
around the particle 
(that is, the part of ${\bf g}$ that diverges as $\mu/R$) 
is correctly matched 
to the corresponding singular behavior of the black hole metric. 
In order to match 
the regular part of the linear metric perturbation 
with the corresponding part of the black hole metric, 
it is necessary to account 
for the motion of the black hole 
with respect to the local inertia frame of the background metric. 
The result is the regular equation of motion. 
This suggests that the stress-energy tensor 
to the second order perturbation can be written as 
\begin{eqnarray}
T^{[1]\mu\nu}+T^{[2]\mu\nu} = 
\mu \int d\tau \tilde v^\mu \tilde v^\nu 
{\delta(x-\tilde x(\tau)) \over \sqrt{-|g^{[0]}|}} 
\,, \label{eq:st12} 
\end{eqnarray}
where $\tilde x^\alpha$ and $\tilde v^\alpha$ 
are the position and $4$-velocity 
of the orbit derived in Sec. \ref{sec:lsf}. 

The metric perturbation is then 
\begin{eqnarray}
g^{[1]}_{\alpha\beta}+g^{[2]SF}_{\alpha\beta} 
= 8\pi\mu \int d\lambda \left({d\tau \over d\lambda}\right)
G_{\alpha\beta\,,\mu\nu}\left(x,\tilde x(\lambda)\right)
\tilde v^\mu \tilde v^\nu 
\,. \label{eq:mem12}
\end{eqnarray}
Recall our definitions (\ref{eq:sf4}) and(\ref{eq:sf6}) 
for the orbital coordinates. 
Here, we use the same formal expressions 
(\ref{eq:geo3})-(\ref{eq:geo6}) for the geodesic, 
but we replace ${\cal E}^a$ and $\chi_\alpha$ 
by $\tilde{\cal E}^a$ and $\tilde\chi_\alpha$. 
Thus we can rewrite (\ref{eq:mem12}) 
by replacing ${\cal E}^a$ and $\chi_\alpha$ of (\ref{eq:lmp_x}) 
with $\tilde{\cal E}^a$ and $\tilde\chi_\alpha$ 
\begin{eqnarray}
g^{[1]}_{\alpha\beta} +g^{[2]SF}_{\alpha\beta} 
&=& 8\pi\mu\int d\lambda \sum_{\omega,k,l,m} 
e^{-i\omega t+im\phi} 
\tilde h^{\omega,k,l,m}_{\alpha\beta}(r,\theta;\tilde{\cal E}) 
e^{i\omega\tilde\chi_t -im\tilde\chi_\phi
-ik\tilde\chi_r-il\tilde\chi_\theta} 
\,, \label{eq:met12_x}
\end{eqnarray}
where $\tilde h^{\omega,k,l,m}_{\alpha\beta}$ 
is defined in the same way as $h^{\omega,k,l,m}_{\alpha\beta}$, 
but with ${\cal E}^a$ replaced by $\tilde {\cal E}^a$. 

We now discuss the effect of the self-force 
on (\ref{eq:met12_x}) by perturbation. 
There are two kinds of effects. 
One is from the dependence of the expansion coefficients 
$\tilde h^{\omega,k,l,m}_{\alpha\beta}$ 
of $\tilde{\cal E}^a$, 
and the other is from the phase function $\tilde\chi_\alpha$ 
in the exponentials. 

We first look at the effect of the expansion coefficients 
$\tilde h^{\omega,k,l,m}_{\alpha\beta}$. 
Using (\ref{eq:sf2}) and (\ref{eq:sf3}), 
the expansion coefficients become 
\begin{eqnarray}
\tilde h^{\omega,k,l,m}_{\alpha\beta}(r,\theta;\tilde{\cal E}) 
\to \tilde h^{\omega,k,l,m}_{\alpha\beta}(r,\theta;{\cal E}_0) 
+{d \over d\tilde{\cal E}^a}
\tilde h^{\omega,k,l,m}_{\alpha\beta}(r,\theta;{\cal E}_0) 
\left<\dot {\cal E}^a\right>\lambda
\,, 
\end{eqnarray}
where we have ignored the oscillating part of $\tilde{\cal E}^a$. 
(The metric perturbation 
due to the oscillating part of $\tilde{\cal E}^a$ 
can be written in the form (\ref{eq:lmp}), 
and does not show any new characteristic feature.) 
One can see 
that the expansion coefficients grow linearly with $\lambda$. 
This is because of the linear growth of ${\cal E}^a$ 
caused by the self-force. 
This leads to 
the linearly growing/decaying feature of the wave amplitude 
through the ${\cal E}^a$-dependence 
of $h^{\omega_{(k,l,m)},k,l,m}_{\alpha\beta}$ as\footnote{
The actual procedure of integration is 
\begin{eqnarray}
\int d\omega e^{-i\omega t}\int dt' t' e^{i(\omega-\omega_0)t'} 
&=&\int d\omega e^{-i\omega t}(-i){d \over d\omega}
\int dt' e^{i(\omega-\omega_0)t'} 
=(2\pi)\int d\omega e^{-i\omega t}(-i){d \over d\omega}
\delta(\omega-\omega_0) 
\nonumber \\ && 
=(2\pi)\int d\omega 
\delta(\omega-\omega_0)i{d \over d\omega}e^{-i\omega t}
=(2\pi)t e^{-i\omega_0 t}
\,.
\end{eqnarray}
} 
\begin{eqnarray}
{2\pi \over \Upsilon_{t(0)}} 8\pi\mu 
\sum_{k,l,m} e^{-i\omega_{(k,l,m)} t+im\phi} 
\left(h^{\omega_{(k,l,m)},k,l,m}_{\alpha\beta}(r,\theta) 
+\dot h^{\omega_{(k,l,m)},k,l,m}_{\alpha\beta}(r,\theta) 
t\right)
e^{i(k\omega_r t^r+l\omega_\theta t^\theta
m\omega_\phi t^\phi)} 
\,, \label{eq:met12_a} 
\end{eqnarray}
where $\dot h^{\omega_{(k,l,m)},k,l,m}_{\alpha\beta}
=(d\tilde h^{\omega,k,l,m}_{\alpha\beta}/d{\cal E}^a) 
\left<\dot {\cal E}^a\right>/\Upsilon_t$. 
This formal expression for the waveform has a new feature, 
that is, the wave amplitude 
changes on the radiation reaction timescale. 

We next see the effect of $\tilde\chi_\alpha$ 
in the exponential of (\ref{eq:met12_x}). 
Using (\ref{eq:sf9}) this exponential has 
a quadratically growing feature 
\begin{eqnarray}
e^{i\omega\tilde\chi_t -im\tilde\chi_\phi
-ik\tilde\chi_r-il\tilde\chi_\theta} 
&\to& 
e^{i\omega\tilde\Upsilon_{t(0)}\lambda
-im\tilde\Upsilon_{\phi(0)}\lambda
-ik\tilde\Upsilon_{r(0)}\lambda
-il\tilde\Upsilon_{\theta(0)}\lambda}
e^{-i\omega\Upsilon_{t(0)}\lambda^t
+im\Upsilon_{\phi(0)}\lambda^\phi
+ik\Upsilon_{r(0)}\lambda^r
+il\Upsilon_{\theta(0)}\lambda^\theta}
\nonumber \\ && 
\times \left\{1
+\left(i\omega\left<\dot \Upsilon_t\right>
-im\left<\dot\Upsilon_\phi\right>
-ik\left<\dot \Upsilon_r\right>
-il\left<\dot\Upsilon_\theta\right>\right)
{\lambda^2 \over 2}\right\}
\,, 
\end{eqnarray}
where we have ignored 
the oscillating part of $\tilde\chi_\alpha$ 
(Again, the metric perturbation 
due to the oscillating part of $\tilde\chi_\alpha$ 
can be written in the form (\ref{eq:lmp}), 
and does not show any new characteristic feature.) 
and we use the renormalized frequencies 
$\tilde\Upsilon_{\alpha(0)}=
\Upsilon_{\alpha(0)}+\Upsilon_{\alpha(0,0)}$. 
We obtain the formal expression as \footnote{
The actual procedure of integration is 
\begin{eqnarray}
\int d\omega e^{-i\omega t}\int dt' t'^2 e^{i(\omega-\omega_0)t'} 
&=&\int d\omega e^{-i\omega t}(-1){d^2 \over d\omega^2}
\int dt' e^{i(\omega-\omega_0)t'} 
=(2\pi)\int d\omega e^{-i\omega t}(-1){d^2 \over d\omega^2}
\delta(\omega-\omega_0) 
\nonumber \\ && 
=(2\pi)\int d\omega 
\delta(\omega-\omega_0)(-1){d^2 \over d\omega^2}e^{-i\omega t}
=(2\pi)t^2 e^{-i\omega_0 t}
\,.
\end{eqnarray}
} 
\begin{eqnarray}
&& {2\pi \over \Upsilon_{t(0)}} 8\pi\mu \sum_{k,l,m} 
e^{-i(\tilde\omega_{(k,l,m)} t+\dot\omega_{(k,l,m)} t^2/2)+im\phi} 
h^{\tilde\omega_{(k,l,m)},k,l,m}_{\alpha\beta}(r,\theta) 
e^{i(k\omega_r t^r+l\omega_\theta t^\theta
m\omega_\phi t^\phi)} 
\,, \label{eq:met12_p} 
\end{eqnarray}
where we used 
\begin{eqnarray}
\tilde\omega_{(k,l,m)} &=& 
m{\tilde\Upsilon_{\phi(0)} \over \tilde\Upsilon_{t(0)}}
+k{\tilde\Upsilon_{r(0)} \over \tilde\Upsilon_{t(0)}}
+l{\tilde\Upsilon_{\theta(0)} \over \tilde\Upsilon_{t(0)}}
\,, \\ 
\dot\omega_{(k,l,m)} &=& 
m\left({\left<\dot \Upsilon_\phi\right> \over \Upsilon_{t(0)}^2}
-{\Upsilon_{\phi(0)}\left<\dot \Upsilon_t\right> \over \Upsilon_{t(0)}^3}\right)
+k\left({\left<\dot \Upsilon_r\right> \over \Upsilon_{t(0)}^2}
-{\Upsilon_{r(0)}\left<\dot \Upsilon_t\right> \over \Upsilon_{t(0)}^3}\right)
\nonumber \\ && 
+l\left({\left<\dot \Upsilon_\theta\right> \over \Upsilon_{t(0)}^2}
-{\Upsilon_{\theta(0)}\left<\dot \Upsilon_t\right> \over \Upsilon_{t(0)}^3}\right)
\,. 
\end{eqnarray}
From the time-dependence of the exponent, 
one can see two key features of the wave phase. 
First, the quadratic growth of the wave phase on the dephasing time 
due to the linear growth of the fundamental frequencies. 
Second, the linear growth of the wave phase 
over the radiation reaction time 
due to the renormalization of the fundamental frequencies. 
Both features have been seen from the orbital evolution 
due to the self-force as we argued in Sec. \ref{sec:lsf}. 

The results of  (\ref{eq:met12_a}) and (\ref{eq:met12_p}) 
suggest that the gravitational field might be well-approximated 
by the form 
\begin{eqnarray}
\mu \sum_{k,l,m} 
h^{k,l,m}_{\alpha\beta}\left(r,\theta;\tilde{\cal E}(t)\right) 
e^{-ik\tilde\chi_r(t)-il\tilde\chi_\theta(t)-im\tilde\chi_\phi(t)}
\,, \label{eq:met12_ap} 
\end{eqnarray}
where $\tilde{\cal E}$ and $\tilde\chi_i (i=r,\theta,\phi)$ 
are functions of the Boyer-Lindquist coordinate time $t$ 
and the radiation reaction effect is included 
in the waveform through those functions.

%%%%%%%%%%%%%%%%%%%%%%%%%%%%%%%%%%%%%%%%%%%%%%%%%%%%%%%%%%%%
\section{Conclusion}
\label{sec:con} 
%%%%%%%%%%%%%%%%%%%%%%%%%%%%%%%%%%%%%%%%%%%%%%%%%%%%%%%%%%%%

In this paper, 
we discuss the modulation of the gravitational waveform 
due to gravitational radiation reaction. 
Since the linear metric perturbation is induced 
only by the background geodesic, 
it is necessary to calculate 
the second order metric perturbation 
in order to see the radiation reaction effect. 
A more rigorous calculation of the second order metric perturbation 
is not yet available. 
For this reason, 
we have considered only a qualitative study, 
extending the technique of formal calculation 
used to obtain the linear metric perturbation in Ref. \cite{rad}. 
The advantage of this technique is 
that one can grasp some key features 
of the gravitational waveform without a complicated calculation. 

There are two kinds of source terms 
for the second order metric perturbation. 
One is the non-linear term of the linear metric perturbation 
$G^{[2]\mu\nu}[{\bf g}^{(1)},{\bf g}^{(1)}]$. 
Because the formal calculation of this  term is triperiodic, 
the part of the second order metric perturbation (\ref{eq:2mpNL0}) 
is also triperiodic. 
The waveform induced by this part 
has exactly the same spectral form 
as that of the linear metric perturbation. 
This part simply changes the amplitude of the waves, 
therefore, describing the correction to the wave propagation 
due to the non-linearity of the Einstein equation. 
The other kind of source term is the point source 
due to the orbital deviation from the background geodesic, 
$T^{[2]\mu\nu}$. 
The part of the the second order metric perturbation 
induced by this source has two new features; 
the linear growth of the wave amplitude 
over the radiation reaction time scale 
[See (\ref{eq:met12_a}).] 
and the quadratic growth of the wave phase 
over the dephasing time scale 
[See (\ref{eq:met12_p}).]. 
From the derivations of (\ref{eq:met12_a}) and (\ref{eq:met12_p}), 
it should be clear that these features reflect 
the secular effect of the orbital evolution due to the self-force 
discussed in Sec. \ref{sec:lsf}. 

These results suggest 
that the effect of radiation reaction on the waveform 
are included in templates of the form (\ref{eq:met12_ap}), 
which is the linear metric perturbation 
with the geodesic constants replaced 
by those which evolve adiabatically due to the self-force. 
This result agrees 
with the procedure of the adiabatic approximation. 
However, strictly speaking 
the perturbation scheme is valid over a dephasing time \cite{adi}, 
and over that time scale, 
only the feature of the quadratic growth of the wave phase 
can be identified as a secular effect. 
Beyond this time scale, it is necessary 
to account for the secular effect of the self-force 
due to the second order metric perturbation. 
One can only discuss those linearly growing features 
in a gauge invariant way 
after calculating the third order metric perturbation. 

Because the perturbation scheme we use 
is limited to the dephasing time scale, 
it is not an appropriate method for calculating 
gravitational waveforms. 
In order to solve this problem, 
we propose an adiabatic expansion 
that is a systematic perturbation 
of a field theory coupled to a particle \cite{adi}. 
Because this expansion recovers 
the adiabatic approximation at leading order, 
our result here supports 
the validity of this new expansion method. 
The application of this technique 
to the adiabatic expansion shall be discussed elsewhere.

%%%%%%%%%%%%%%%%%%%%%%%%%%%%%%%%%%%%%%%%%%%%%%%%%%%%%%%%%%%%
\section*{Acknowledgement} 
%%%%%%%%%%%%%%%%%%%%%%%%%%%%%%%%%%%%%%%%%%%%%%%%%%%%%%%%%%%%

We thank Prof. Richard Price for fruitful discussion. 
We thank Dr. Steve Drasco for carefully reading the material. 
This work is supported 
by NSF grant PHY-0601459, NASA grant NNX07AH06G, 
NNG04GK98G and the Brinson Foundation.

\appendix
%%%%%%%%%%%%%%%%%%%%%%%%%%%%%%%%%%%%%%%%%%%%%%%%%%%%%%%%%%%%
\section{Formal Expression of the Linear Metric Perturbation}
\label{app:lin} 
%%%%%%%%%%%%%%%%%%%%%%%%%%%%%%%%%%%%%%%%%%%%%%%%%%%%%%%%%%%%

In this appendix, we review 
the formal calculation of the linear metric perturbation 
given as (\ref{eq:met}). 
With the ansatz (\ref{eq:tgr}), 
we decompose the Green's function as 
\begin{eqnarray}
G_{\alpha\beta\,\mu\nu}(x,x') 
= \sum_{\omega,m} 
g^{\omega,m}_{\alpha\beta\,\mu\nu}(r,r';\theta,\theta') 
e^{-i\omega(t-t')+im(\phi-\phi')}. 
\end{eqnarray}
We plug this into (\ref{eq:met}) and we have 
\begin{eqnarray}
g^{[1]}_{\alpha\beta}(x) 
= 8\pi\mu\int d\lambda \left({d\tau \over d\lambda}\right) 
\sum_{\omega,m}
g^{\omega,m}_{\alpha\beta\,\mu\nu}(r,\bar r;\theta,\bar\theta) 
v^\mu v^\nu
e^{-i\omega(t-\bar t)+im(\phi-\bar\phi)}. 
\,. \label{eq:met_x} 
\end{eqnarray}
At this point, we recall the periodic feature of the orbit 
(\ref{eq:geo3})-(\ref{eq:geo6}). 
It is easy to see that 
$d\tau/d\lambda$ and 
$g^{\omega,m}_{\alpha\beta\,\mu\nu}(r,\bar r;\theta,\bar\theta)$ 
can be expanded as a discrete Fourier series 
in $e^{-ik\chi_r-il\chi_\theta}$. 
From (\ref{eq:geo1}) and (\ref{eq:geo2}), 
$v^\mu v^\nu$ can be similarly expanded. 
As for $e^{i\omega\bar t -im\bar\phi}$, 
one may expand by the same discrete Fourier series 
except the factor $e^{i\omega\chi_t -im\chi_\phi}$. 
In summary, one can formally rewrite (\ref{eq:met_x}) as 
\begin{eqnarray}
g^{[1]}_{\alpha\beta}(x) 
&=& 8\pi\mu\int d\lambda \sum_{\omega,k,l,m} 
e^{-i\omega t+im\phi} 
h^{\omega,k,l,m}_{\alpha\beta}(r,\theta) 
e^{i\omega\chi_t -im\chi_\phi-ik\chi_r-il\chi_\theta} 
\,. \label{eq:lmp_x}
\end{eqnarray}
By integrating over $\lambda$, we obtain 
\begin{eqnarray}
g^{[1]}_{\alpha\beta}(x) 
&=& {2\pi \over \Upsilon_t} 8\pi\mu \sum_{k,l,m} 
e^{-i\omega_{(k,l,m)} t+im\phi} 
h^{\omega_{(k,l,m)},k,l,m}_{\alpha\beta}(r,\theta) 
e^{im\Upsilon_\phi(\lambda^\phi-\lambda^t)
+ik\Upsilon_r(\lambda^r-\lambda^t)
+il\Upsilon_\theta(\lambda^\theta-\lambda^t)} 
\,, 
\end{eqnarray}
which is equal to (\ref{eq:lmp}).

%%%%%%%%%%%%%%%%%%%%%%%%%%%%%%%%%%%%%%%%%%%%%%%%%%%%%%%%%%%%
\section{Regularization of the ultraviolet divergence}
\label{app:reg1} 
%%%%%%%%%%%%%%%%%%%%%%%%%%%%%%%%%%%%%%%%%%%%%%%%%%%%%%%%%%%%

We use the regularization prescription 
for the ultraviolet divergence proposed in Ref. \cite{2nd}. 
By the consistency of the matched asymptotic expansion, 
we know that the divergence behavior 
of the second order metric perturbation becomes 
${\bf g}^{[2]NL} \propto \mu^2/R^2$, 
with respect to the local inertial coordinates 
in the neighborhood of the geodesic. 
($R$ is the spatial distance from the particle 
in the local inertia frame.) 
The idea in Ref. \cite{2nd} is to subtract this divergence 
by using the quadratic combination 
of the scalar field induced by a point particle 
and calculate the remaining regular part by the Green's method. 
We use the scalar field that satisfies 
\begin{eqnarray}
\Box \Phi &=& \int d\tau 
{\delta(x-\bar x(\tau)) \over \sqrt{-|g^{[0]}|}} \,, \label{eq:sce}
\end{eqnarray}
under the retarded boundary condition, 
where $\bar x^\mu(\tau)$ is the same geodesic 
used in (\ref{eq:st1}) for the linear metric perturbation. 
The resulting scalar field has the divergence behavior 
$\Phi \propto 1/R$ near the geodesic. 
With appropriate tensors 
$k^{[2]}_{\alpha\beta}$ and $k^{[1]}_{\alpha\beta}$, 
which are regular around the orbit, 
one may construct 
\begin{eqnarray}
g^{(S)[2]NL}_{\alpha\beta} = 
k^{[2]}_{\alpha\beta}\Phi^2+k^{[1]}_{\alpha\beta}\Phi 
\,, \label{eq:2mpNLS} 
\end{eqnarray}
such that ${\bf g}^{[2]NL}$ and ${\bf g}^{(S)[2]NL}$ 
have the same singular behavior near the geodesic. 
We define the remaining part by 
\begin{eqnarray}
g^{(R)[2]NL}_{\alpha\beta} &=& 
g^{[2]NL}_{\alpha\beta} -g^{(S)[2]NL}_{\alpha\beta} 
\,, 
\end{eqnarray}
which satisfies 
\begin{eqnarray}
G^{[1]\mu\nu}[{\bf g}^{(R)(2)NL}] &=& 
-G^{[1]\mu\nu}[{\bf g}^{(S)(2)NL}] 
-G^{[2]\mu\nu}[{\bf g}^{(1)},{\bf g}^{(1)}] 
\,. \label{eq:2einNL} 
\end{eqnarray}
Because $g^{(R)[2]NL}_{\alpha\beta}$ is regular, 
the RHS of (\ref{eq:2einNL}) must be regular 
near the geodesic, 
and ${\bf g}^{(R)(2)NL}$ can be written as 
\begin{eqnarray}
g^{(R)[2]NL}_{\alpha\beta}(x) &=& -\int_{-\infty<r^*<\infty} 
d^4 x' G_{\alpha\beta\, \mu\nu}
\left(G^{[1]\mu\nu}[{\bf g}^{(S)(2)NL}] 
+G^{[2]\mu\nu}[{\bf g}^{(1)},{\bf g}^{(1)}]\right) 
\,. \label{eq:2mpNLR} 
\end{eqnarray}

In order to see the formal expression for (\ref{eq:2mpNLR}), 
it is necessary to know the formal expression for (\ref{eq:2mpNLS}). 
The expression for $\Phi$ can be obtained 
in a similar manner as (\ref{eq:lmp}). 
Because the scalar Green's function of (\ref{eq:sce}) 
can be written in the form 
\begin{eqnarray}
G(x,x') = G(t-t',\phi-\phi';r,r',\theta,\theta') 
\,, \label{eq:sgr} 
\end{eqnarray}
similarly to (\ref{eq:tgr}), 
the formal expression of $\Phi$ becomes 
\begin{eqnarray}
\Phi(x) &=& \sum_{k,l,m}
e^{-i \omega_{(k,l,m)} t+i m \phi}
\Phi_{(k,l,m)}(r,\theta;{\cal E}^a)
e^{i (k \omega_r t^r +l \omega_\theta t^\theta 
+m \omega_\phi t^\phi)} 
\,. \label{eq:scl} 
\end{eqnarray}
The structure of the singularity depends 
only on the local background geometry, 
and the components of $k^{[1]}_{\alpha\beta}$ 
and $k^{[2]}_{\alpha\beta}$ can be constructed 
only from the background curvature along the orbit. 
Therefore, we have 
\begin{eqnarray}
k^{[n]}_{\alpha\beta} &=& \sum_{k,l,m}
e^{-i \omega_{(k,l,m)} t+i m \phi}
k^{[n]}_{\alpha\beta(k,l,m)}(r,\theta;{\cal E}^a)
e^{i (k \omega_r t^r +l \omega_\theta t^\theta 
+m \omega_\phi t^\phi)} \,, 
\end{eqnarray}
for $n=1,2$. 
In summary, the RHS of (\ref{eq:2einNL}) has 
the same formal structure as (\ref{eq:2ein0}).

%%%%%%%%%%%%%%%%%%%%%%%%%%%%%%%%%%%%%%%%%%%%%%%%%%%%%%%%%%%%
\section{Regularization of the infrared divergence}
\label{app:reg2} 
%%%%%%%%%%%%%%%%%%%%%%%%%%%%%%%%%%%%%%%%%%%%%%%%%%%%%%%%%%%%

Though we can remove the ultraviolet divergence 
by calculating (\ref{eq:2mpNLR}) 
instead of calculating (\ref{eq:2mpNL}), 
it does not guarantee that we can also remove 
the infrared divergence, 
i.e. the integration over the spatial volume of (\ref{eq:2mpNLR}) 
becomes divergent 
at the large spatial radius and at the black hole horizon. 

A formal method of avoiding this divergence 
is to introduce the radial regulator $\Gamma$ 
to (\ref{eq:2mpNLR}) as 
\begin{eqnarray}
g^{(R)[2]NL}_{\alpha\beta}(x;\Gamma) &=& 
-\int_{-\Gamma<r^*<\Gamma} d x'^4 \,
G_{\alpha\beta\mu\nu}
\left(G^{[1]\mu\nu}[{\bf g}^{(S)(2)NL}] 
+G^{[2]\mu\nu}[{\bf g}^{(1)},{\bf g}^{(1)}]\right) 
\,. \label{eq:2mpNLRe} 
\end{eqnarray}
By definition, 
${\bf g}^{(R)[2]NL}(x;\Gamma)$ satisfies (\ref{eq:2einNL}) 
for $-\Gamma<r^*<\Gamma$, 
and it diverges as 
${\bf g}^{(R)[2]NL}(x;\Gamma) \propto \ln(\Gamma)$, 
when $\Gamma \to \infty$. 

The counter term to cancel this divergence 
${\bf g}^{(C)[2]NL}(x;\Gamma)$ 
must satisfy the vacuum linearized Einstein equation 
so that ${\bf g}^{(R)[2]NL}+{\bf g}^{(C)[2]NL}$
still satisfies (\ref{eq:2einNL}). 
Because ${\bf g}^{[2]NL}$ is intrinsically divergent 
along the orbit, 
one may consider a point source moving along the geodesic 
which induces ${\bf g}^{(C)[2]NL}(x;\Gamma)$ as 
\begin{eqnarray}
T^{[2]NL\,\mu\nu}(x;\Gamma) &=& \ln(\Gamma) 
\mu^2\int d\tau \, 
{\hat t}^{[2]\mu\nu}{\delta(x-\bar x(\tau)) \over \sqrt{-|g^{[1]}|}} 
\,, \label{eq:stc} 
\end{eqnarray}
where $\bar x^\alpha(\tau)$ 
is the same geodesic as in (\ref{eq:st1}). 
We have the counter term 
\begin{eqnarray}
g^{(C)[2]NL}_{\alpha\beta}(x;\Gamma) 
&=& 8\pi \int d x'^4\, G_{\alpha\beta\mu\nu}(x,x')
T^{[2]NL\,\mu\nu}(x;\Gamma) 
\nonumber \\ 
&=& 8\pi\mu^2 \ln(\Gamma)\int d\tau 
{\hat t}^{[2]\mu\nu}G_{\alpha\beta\mu\nu}(x,\bar x)
\,. 
\end{eqnarray} 
Here ${\hat t}^{[2]\mu\nu}$ is the tensor derivative operator 
defined along the orbit such that 
$\left({\bf g}^{(R)[2]NL}+{\bf g}^{(C)[2]NL}\right)(x;\Gamma)$ 
becomes finite everywhere except along the geodesic 
when $\Gamma \to \infty$. 

We discuss the formal expression of $\lim_{\Gamma \to\infty}
\left({\bf g}^{(R)[2]NL}+{\bf g}^{(C)[2]NL}\right)(x;\Gamma)$ 
following this regularization prescription. 
For a finite $\Gamma$, as we may see from (\ref{eq:2mpNLRe}), 
${\bf g}^{(R)[2]NL}(x;\Gamma)$ has 
the same formal expression as (\ref{eq:2mpNL0}) 
because the source term of (\ref{eq:2mpNL0}) 
has the same formal expression as (\ref{eq:2ein0}). 
By ${\bf g}^{(C)[2]NL}(x;\Gamma)$, 
we simplify subtract the linear divergent piece 
of ${\bf g}^{(R)[2]NL}(x;\Gamma)$, 
i.e. we have 
\begin{eqnarray}
{\bf g}^{(C)[2]NL}(x;\Gamma) 
= -\ln(\Gamma) \lim_{\Gamma' \to \infty}
{{\bf g}^{(R)[2]NL}(x;\Gamma') \over \ln(\Gamma')} \,, 
\end{eqnarray} 
and this operation does not alter the formal expression. 
Thus, by the infrared regularization, 
the part of the second order metric perturbation 
due to the non-linear term of (\ref{eq:2ein}) 
has the formal expression of (\ref{eq:2mpNL0}).

%%%%%%%%%%%%%%%%%%%%%%%%%%%%%%%%%%%%%%%%%%%%%%%%%%%%%%%%%%%

%%%%%%%%%%%%%%%%%%%%%%%%%%%%%%%%%%%%%%%%%%%%%%%%%%%%%%%%%%%

\end{document}